\documentclass[USenglish]{article}	
\usepackage[utf8]{inputenc}				%(only for the pdftex engine)
\usepackage[big,online]{dgruyter}	%values: small,big | online,print,work
\usepackage{lmodern} 
\usepackage{microtype}
\usepackage[numbers,square,sort&compress]{natbib}
\usepackage{color}
\theoremstyle{dgthm}

\theoremstyle{dgdef}

\begin{document}

%%%--------------------------------------------%%%
	\articletype{Research Article \hfill Accepted version of Nanophotonics 10, 2717-2728 (2021)}
%	\received{Month	DD, YYYY}
%	\revised{Month	DD, YYYY}
%  \accepted{Month	DD, YYYY}
 % \journalname{De~Gruyter~Journal}
 % \journalyear{YYYY}
 % \journalvolume{XX}
 % \journalissue{X}
  \startpage{1}
%  \aop
 % \DOI{10.1515/sample-YYYY-XXXX}
%%%--------------------------------------------%%%

\title{Local field effects in ultrafast light-matter interaction measured by pump-probe spectroscopy of monolayer MoSe$_{\boldsymbol 2}$}
%Exciton dynamics in hBN/MoSe$_{\boldsymbol 2}$/hBN heterostructures measured in pump-probe spectroscopy}
\runningtitle{Exciton dynamics in MoSe$_{2}$}
%\subtitle{(Supporting Information)}

\author*[1]{Aleksander~Rodek}
%\ use * to mark the author as the corresponding author
\author[2,3]{Thilo~Hahn}
\author[1,4]{Jacek~Kasprzak}
\author[1]{Tomasz~Kazimierczuk}
\author[1]{Karol~Nogajewski}
\author[1]{Karolina~Po\l{}czy\'nska}
\author[5]{Kenji~Watanabe}
\author[5]{Takashi~Taniguchi}
\author[2]{Tilmann~Kuhn}
\author[3]{Pawe\l{}~Machnikowski}
\author[1,6]{Marek~Potemski}
\author[3]{Daniel~Wigger}
\author[1]{Piotr~Kossacki}
 
\runningauthor{A.~Rodek, T. Hahn et al.}
 
%\runningauthor{F.~Author et al.}

\affil[1]{\protect\raggedright 
Institute of Experimental Physics, Faculty of Physics, University of Warsaw, 02-093 Warszawa, Poland, e-mail: aleksander.rodek@fuw.edu.pl}
\affil[2]{\protect\raggedright 
Institut f\"ur Festk\"orpertheorie, Universit\"at M\"unster, 48149 M\"unster, Germany, e-mail: t.hahn@wwu.de}
\affil[3]{\protect\raggedright 
Department of Theoretical Physics, Wroc{\l}aw University of Science and Technology, 50-370 Wroc{\l}aw, Poland}
\affil[4]{\protect\raggedright 
Universit\'{e} Grenoble Alpes, CNRS, Grenoble INP, Institut N\'{e}el, 38000 Grenoble, France}
\affil[5]{\protect\raggedright 
National Institute for Materials Science, Tsukuba, Ibaraki 305-0044, Japan}
\affil[6]{\protect\raggedright 
Laboratoire National des Champs Magn\'{e}tiques Intenses, CNRS-UGA-UPS-INSA-EMFL, 38042 Grenoble, France}

%\communicated{...}
%\dedication{...}
	
\abstract{Using a novel approach to ultrafast resonant pump-probe spectroscopy we investigate the spectral shape and dynamics of absorption features related to the A exciton in an hBN/MoSe$_2$/hBN van der Waals heterostructure. While in a pure two-level system a pump-probe experiment measures the occupation or the polarization dynamics, depending on the time ordering of the pulse pair, in the transition metal dichalcogenide (TMD) system both quantities get thoroughly mixed by strong exciton-exciton interaction. We find that for short positive delays the spectral lines experience pronounced changes in their shape and energy and they relax to the original situation on a picosecond time scale. For negative delays distinctive spectral oscillations appear indicating the first-time observation of perturbed free induction decay for a TMD system. The comparison between co-circular and cross-circular excitation schemes further allows us to investigate the rapid inter-valley scattering. By considering a three-level system as a minimal model including the local field effect, excitation induced dephasing and scattering between the excited states we explain all phenomena observed in the experiment with excellent consistency. Our handy model can be even further reduced to two levels in the case of a co-circular excitation, for which we derive analytic expressions to describe the detected signals. This allows us to trace back the spectral shapes and shifts to the impact of local field effect and excitation induced dephasing thus fully reproducing the complex behavior of the observed effects.}


\maketitle

%%%MAIN TEXT%%%%
\section{Introduction}
%%%%%%%%%%%%%%
Absorption and emission of light from semiconducting transition metal dichalcogenide (TMD) monolayers (MLs) is governed by tightly bound excitons~\cite{KoperskiNanoPho2016, WangRMP18}. The surface of a TMD single layer can be protected by embedding it between sheets of high quality, atomically flat hexagonal boron nitride (hBN). Such shielding prevents from aging effects due to erosive chemistry at ambient conditions~\cite{CadizPRX17, Ajayi2DMater17}. Moreover, the hetero-structuring can flatten the TMD monolayers, eradicating disorder due to wrinkling and strain variations across macroscopic distances of many microns~\cite{JakubczykACSNano19, BoulePRM20}. As a result, exciton transitions in modern hBN/TMD/hBN heterostructures reach line widths close to the homogeneous limit in the range of several milli-electronvolts (meV) at cryogenic temperatures. Such energies correspond to exciton dynamics on the sub-picosecond time scale as confirmed by nonlinear spectroscopy~\cite{JakubczykNanoLett16}. Femto-second (fs) multi-pulse spectroscopy is therefore required in order to investigate exciton dynamics in such systems.

Here, we perform resonant fs pump-probe measurements on the hBN/MoSe$_2$(ML)/hBN heterostructure sample shown in Fig.~\ref{fig:sample}(a). Considering co- and cross-circularly polarized excitations we study population decay and inter-valley scattering rates. We analyze the shape of the absorption features close to temporal overlap of the laser pulses in the regime of strong optical excitation. Characteristic energy shifts and line shapes give insight into the impact of local field effects and excitation induced dephasing (EID). Importantly, when probing the coherence by a probe-pump sequence, we detect characteristic spectral oscillations on the exciton line reaching over few tens of meV. This is usually attributed to the extinction of the exciton polarization transient by the second pulse~\cite{GuentherPRL02, FrasNatPhot16, MondalJCM18} or by EID~\cite{KochJPC88, LindbergJOSAB88}. Here, it is additionally related to a rapid change of the character of the pump-probe signal accompanied by a frequency shift introduced by the local field effect~\cite{HahnNJP21}. This study is the first reported observation of this effect for excitons in TMDs. Our data are simulated by a few-level model including local field effects and EID, reaching excellent agreement with the experiment.

%%%%%%%%
\section{Sample and experiment}
%%%%
\begin{figure}[tb]
	\centering
	\includegraphics[width=0.6\columnwidth]{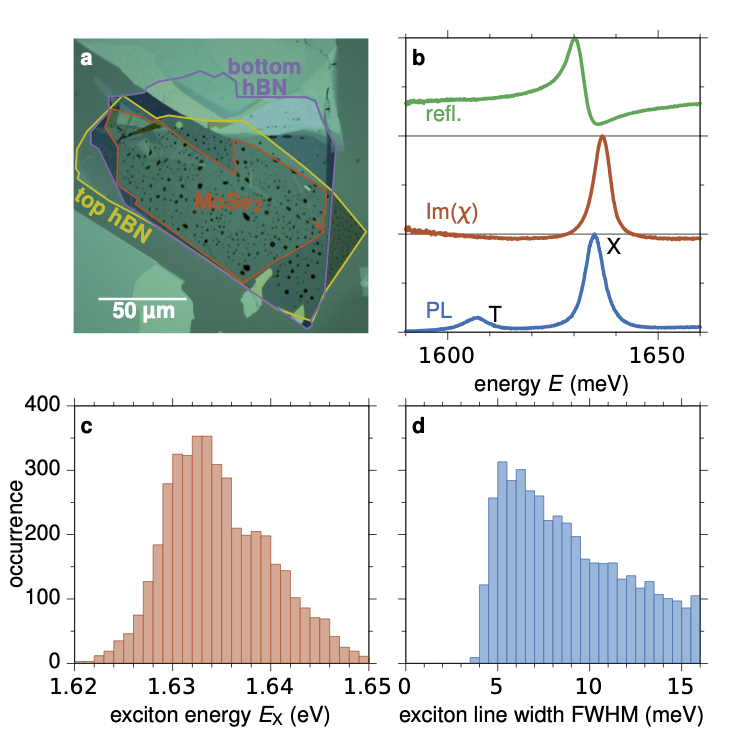}
	\caption{Sample characterization. (a)~White light image of the investigated heterostructure. The bottom hBN is marked in violet, the MoSe$_2$ ML in orange and the top hBN in yellow. (b)~Typical normalized micro-reflectance (top), the corresponding susceptibility Im($\chi$) (center), and micro-photoluminescense (PL) spectra (bottom) at 5~K, measured around the fundamental exciton transition (X). The trion is marked by T. (c) and (d)~Statistics of exciton center transition energy and linewidth, respectively, illustrating inhomogeneities across the sample.}
	\label{fig:sample}
\end{figure}
%%%%
Our hBN/MoSe$_2$(ML)/hBN heterostructure (hBN thickness $d_{\rm bottom}=85$~nm and $d_{\rm top}=20$~nm) was fabricated using a triaxial micromanipulator by a PDMS dry transfer method~\cite{Castellanos_Gomez2DMater14}. The used layers were exfoliated from bulk crystals with a standard micro-mechanical exfoliation technique. The sample was characterized by micro-photoluminescence (micro-PL) mapping at cryogenic temperature of $T=5$~K with continuous wave excitation at $\lambda=630$~nm. The measured spectra in Fig.~\ref{fig:sample}(b) display all characteristic features for this type of heterostructure found in the literature~\cite{Ajayi2DMater17}. The PL in the bottom panel consist of a pronounced neutral exciton line (X) and a charged exciton transition (trion,~T) shifted to lower energies by $\Delta_{\rm XT}\approx 30$~meV. Statistical analysis of the PL-map was performed by fitting the neutral exciton line with a standard Gaussian spectral shape. The histogram of exciton emission energies in Fig.~\ref{fig:sample}(c) forms a smooth, peaked distribution between 1625~meV and 1650~meV. This effect is caused by the remaining local strain~\cite{BoulePRM20} of the crystal lattice induced in the stamping process as well as changes in the dielectric environment influenced by the presence of air bubbles trapped between the different layers of the heterostructure stack. These inhomogeneities are also reflected by the distribution of the exciton's spectral widths in Fig.~\ref{fig:sample}(d), where we find that only a small fraction of measured spectra show a minimal width around 5~meV. Compared to previous measurements reporting widths of 2~meV~\cite{Ajayi2DMater17, BoulePRM20, JakubczykNanoLett16} we can conclude that inhomogeneous broadening could still contribute to the optical response for the narrowest spectra. The further studies are done on selected spots exhibiting low linewidth broadening (5 meV) and spatial homogeneity as extracted from the PL mapping. In particular, we aim to reduce the influence of inhomogeneous broadening to avoid additional shortening of the exciton dephasing time and to stay within the temporal resolution of our experiment~\cite{Jakubczyk_2018}. %A full characterization of the influence of inhomogeneous broadening on the signal dynamics is beyond the scope of this article.

The applied pump-probe technique is a convenient tool to measure the dynamics of charge carriers and excitons in various semiconductor nanostructures~\cite{RossiRMP02, schafer2013}, that employs a pair of fs laser pulses. Usually, it consists of exciting the sample with a strong optical pulse, called the pump, followed by a weaker pulse, called the probe. The intensity and spectral lineshape of the differential reflection signal launched by the probe are sensitive to the modifications induced in the medium by the pump pulse. Dynamics of such variations are monitored by varying the time interval between pump and probe, called the delay $\tau$. The temporal resolution in our setup is only limited by the duration of the pulses $\Delta t=21$~fs (FWHM 50~fs), measured via intensity autocorrelation (see Supporting Information). The pulses themselves are generated in a 76~MHz oscillator pumped with a Ti:Sapphire laser and are tuned to $\lambda=760$~nm, which corresponds to the neutral exciton transition at a temperature of 5~K.

We apply a novel approach to the pump-probe experiment in a micrometer resolution setup. It is performed in back-reflection geometry with efficient spatial separation between the pump and the probe beam. The laser pulses are focused on the sample with a 4~mm focal length, high-NA aspheric lens with 5 mm total diameter. The diffraction limit of the used laser beams corresponds to a spot diameter of $d=3.8$~\textmu m. After the reflection from the sample the probe beam is directed into an imaging spectrometer and recorded on a CCD camera. Spatial separation between the parallel pump and probe laser beams allows for a high degree of extinction of the pump signal (in the order of $10^4$) on the CCD, while maintaining their overlap on the sample. This makes it possible to perform measurements with spectrally degenerate laser pulses that have the same energy and polarization. Furthermore, this method is a complementary alternative to technically more involved coherent detection via optical heterodyning employed in recent experiments~\cite{JakubczykACSNano19, JakubczykNanoLett16}. The reached micrometer resolution is also important for the case of heterostructures with a significant inhomogeneity across the sample where it allows for isolating spots of good, homogeneous optical quality. While recent developments in epitaxial techniques partly solve this issue~\cite{PacuskiNanoLett20}, still a notable amount of scientific research relies on the preparation of samples by the lift-off method. Finally, small residual interference originating from the cross-talk between the pump and the probe is removed by periodically changing the optical path of the pump beam and averaging the signal over time. The relative change in the optical path is generated by a mirror located on a piezo element that oscillates with a frequency of about 30~Hz. The total spatial shift corresponds approximately to the laser wavelength, thus allowing for filtering out interference without deteriorating the temporal resolution. The experiments are performed at a temperature of $T=5$~K. The reflectance measured at the area investigated in the pump-probe study is shown in Fig.~\ref{fig:sample}(b) (top panel). From this signal we calculate the imaginary part of the optical susceptibility Im($\chi$) via a Kramers-Kronig transform according to Ref.~\cite{RochNatNano19} in the middle panel.

%%%%%%%%
\section{Theory}
\begin{figure}[h]
	\centering
	\includegraphics[width=0.45\columnwidth]{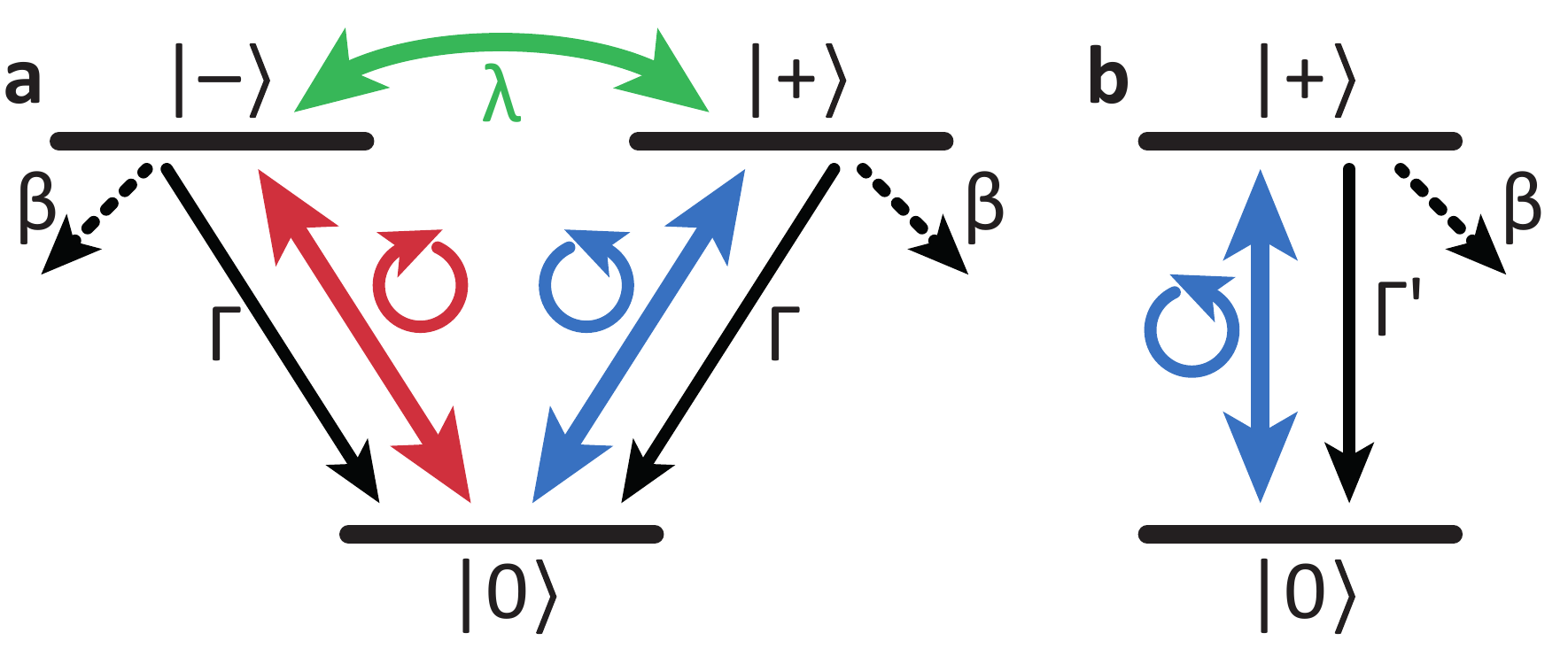}
	\caption{Level structure of the model. (a) Three-level system (3LS) in a V-shape including scattering between the excited states. (b) Reduced two-level system (2LS) considering only one polarization orientation.}
	\label{fig:levels}
\end{figure}

In the experiment we are resonantly exciting the lowest exciton states, i.e., the A excitons in the K and K' valley, with in general differently circularly polarized laser pulses. To treat this system theoretically, a minimum of three states is required as schematically shown in Fig.~\ref{fig:levels}(a). The ground state $\left|0\right>$ has no exciton and the two excited states $\left|\pm\right>$ with the same energy $\hbar\omega_0$ have an exciton in the K or the K' valley, respectively. Each circular polarization orientation addresses one of the two excitons, e.g., $\sigma_-$-polarized light excites the $|-\rangle$ exciton and $\sigma_+$-polarized light the $|+\rangle$ exciton. Further we include scattering between the two excitons, which leads to a transition of the occupation of the excited states. 
%%%

In practice we consider the following equations of motion to describe the microscopic polarizations $p_+=\left<\left| 0\right>\!\left< +\right|\right>$ and $p_-=\left<\left| 0\right>\!\left< -\right|\right>$ of the two excitons and the respective occupations $n_+=\left<\left| +\right>\!\left< +\right|\right>$ and $n_-=\left<\left| -\right>\!\left< -\right|\right>$
%%%
\begin{subequations}\label{eq:3LS}\begin{align}
	\frac{d p_\pm}{d t} &= i(1-2n_\pm)[\Omega_\pm(t) + V p_\pm] \notag\\
	& -  [ \beta + W (n_\pm + n_\mp) + i\omega_0 ] p_\pm \,,\\
	\frac{d n_\pm}{d t} &= 2{\rm Im}[\Omega_\pm^*(t)p_\pm]- \Gamma n_\pm - \lambda (n_\pm - n_\mp)\,.
\end{align}\end{subequations}
%%%
The excitons' decay rate $\Gamma$ and dephasing rate $\beta$ are the same for both excitons, and the transition rate between the exciton occupations is $\lambda$. Especially the phenomenological dephasing rate $\beta$ may include different microscopic sources like fluctuations of the environment or phonon scattering which are for simplicity not further specified in this context. The process of occupation transfer leading to valley depolarization is a known phenomenon in TMD materials~\cite{MakNN2012, ZhuPRB2014, SchmidtNL2016}. While the reason for the occupation transfer is still under debate, we choose here the simplest possible model which is a rate equation balancing the occupations in both valleys. Among the proposed valley depolarization mechanisms are scattering processes with phonons~\cite{JeongNanoscale20} and exciton-exciton interaction~\cite{MahmoodNL2018}. Consistently, we choose for the simulations a scattering rate in the sub-ps range. We always consider $\beta>\Gamma/2+\lambda/2$ to ensure that the density matrix is positive-definite. Note that since the states $\left|+\right>$ and $\left|-\right>$ are only coupled via scattering processes no coherence $\left<\left|+\right>\left<-\right|\right>$ is created. The external optical driving is given by the electric field of the pulses $\boldsymbol{E}_\pm (t)$ translating here to instantaneous Rabi frequencies $\Omega_\pm(t) = \boldsymbol{M}_\pm\cdot \boldsymbol{E}_\pm(t)/\hbar$, where $\boldsymbol M_\pm$ are the dipole moments, allowing to address the two excited states individually. In addition we consider a local field effect parametrized by $V$ and an excitation induced dephasing (EID) by $W$. We want to remark that the local field term results from a combination of different microscopic phenomena~\cite{KatschPRL20}: band gap renormalization, reduction of the exciton binding energy due to Pauli blocking and exciton-exciton interaction. Following the motivation of this effect in Ref.~\cite{wegenerPRA90} the local field coupling should depend on the emission polarization of the considered exciton. Therefore, we only consider the local field coupling within a given valley. The EID stems from exciton-exciton scattering events and depends therefore on the total exciton density~$n_+ + n_-$. Note that here $V$ and $W$ are real-valued, while in Ref.~\cite{HahnNJP21} the two quantities were summed in a complex-valued $V$.

To simulate the experimentally detected pump-probe signal we numerically calculate the system's dynamics for a two-pulse excitation with
%%%
\begin{align}
	\Omega_\pm(t) %&=& \Omega_{\pm,\rm pump}(t) +  \Omega_{\pm,\rm probe}(t) \nonumber\\
			&= \frac{1}{\sqrt{8\pi}\Delta t} \Bigg\{ \theta_1e^{i\phi_1} \exp\left[-\frac{(t+\tau)^2}{2(\Delta t)^2}\right] 
		 + \theta_2 e^{i\phi_2}\exp\left[-\frac{t^2}{2(\Delta t)^2}\right] \Bigg\} e^{-i(\omega_0-V)t} \,,
\end{align}
%%%
where the maximum of the probe pulse defines the time $t=0$. The delay between the pulses is $\tau$ and the pulses have phases $\phi_1$ and $\phi_2$, pulse areas $\theta_1$ and $\theta_2$ and the same duration $\Delta t$. Both pulses are resonant to the exciton transition which is additionally renormalized by the local field strength $V$ in this model~\cite{HahnNJP21}. We identify the pulse with index 1 as the pump pulse and pulse 2 as the probe, meaning $\theta_1\gg \theta_2$ in agreement with the experiment. A positive delay $\tau>0$ indicates that the probe pulse arrives after the pump pulse, while the inverse ordering is described by negative delays. To isolate the pump-probe signal we filter the exciton polarization after the probe pulse $p_2(t,\tau)$ with respect to the phase of the probe pulse, which characterizes the propagation direction $\boldsymbol k_2$ of this pulse in the experiment, via
%%%
\begin{equation}
	p_{\rm  pp}(t,\tau) =  \frac{1}{2\pi}\int\limits_0^{2\pi} p_2(t,\tau) e^{-i\phi_2}\,{\rm d}\phi_2\,.
\end{equation}
%%%
From this we simply retrieve the pump-probe spectrum via~\cite{LindbergJOSAB88,balslevPRB89,HenzlerPRL21}
%%%
\begin{equation}
	S_{\rm pp}(\omega,\tau) \sim \frac{{\rm Im} \left[{\widetilde{E}}^*_{\rm probe}(\omega)\,\widetilde{p}_{\rm pp}(\omega,\tau)\right]}{\left|\widetilde{E}_{\rm probe}(\omega)\right|^2}\,, \label{eq:pp_signal}
\end{equation}
%%%
where the tilde marks the Fourier transformed signal from the time $t$ to the spectral domain, defined as
%%%
\begin{equation}
	\widetilde{p}_{\rm pp} (\omega, \tau) = \mathcal{F} [ p_{\rm pp}(t,\tau)](\omega) = \int\limits_{-\infty}^\infty p_{\rm pp}(t,\tau) e^{i\omega t} \,{\rm d}t\,.
\end{equation}
%%%

%%%%%%%%%
In the following part of this theory section we consider a co-circular excitation which optically addresses only the states $\left|0\right>$ and $\left|+\right>$. In Ref.~\cite{HahnNJP21} we have shown that for a two-level system (2LS) the dynamics can be calculated analytically and brought into a compact form in the limit of ultrafast laser pulses. To carry out an equivalent analysis here we will reduce the three-level system (3LS) to a 2LS. To do so in principle we just have to disregard the exciton scattering rate choosing $\lambda=0$. By doing so we have to take care of the other rates in the model. The scattering from $\left|+\right>$ into $\left|-\right>$ leads to a dynamical loss of population of the optically driven exciton. Therefore, we introduce the effective decay rate~$\Gamma^\prime$ to compensate for the in general time-dependent influence of $\lambda\neq 0$. When comparing the ultrafast pulse limit in the 2LS with finite pulse durations in the 3LS this loss of occupation results in slightly different occupations for the same pulse area directly after the pulse. A detailed discussion of this effect for the present parameters is given in the Supporting Information. The phenomenological dephasing rate $\beta$ is considered to stay the same because it already sums up all dephasing processes in the 3LS. In the following we set $p=p_+$, $n=n_+$, and $\Omega(t) = \Omega_+(t)$ and the reduced equations of motion read
%%%
\begin{subequations}\begin{align}
	\frac{d p}{d t} &= i(1-2n) [\Omega(t) + Vp] - (\beta+Wn + i\omega_0) p  \,,\\
	\frac{d n}{d t} &= 2{\rm  Im}[p\Omega^\ast (t)] - \Gamma^\prime n \,.
\end{align}\end{subequations}
%%%

In the limit of ultrafast laser pulses, treated as delta functions in time, the full system's dynamics can be solved analytically as shown in Ref.~\cite{HahnNJP21}. Based on the analytical expression derived in the following, we can explain the lineshapes and extract approximations for energy shifts and dynamics that we will later compare to our numerical simulations for non-vanishing pulse durations and the experimental findings.

\begin{figure}[h]
	\centering
	\includegraphics[width=0.6\columnwidth]{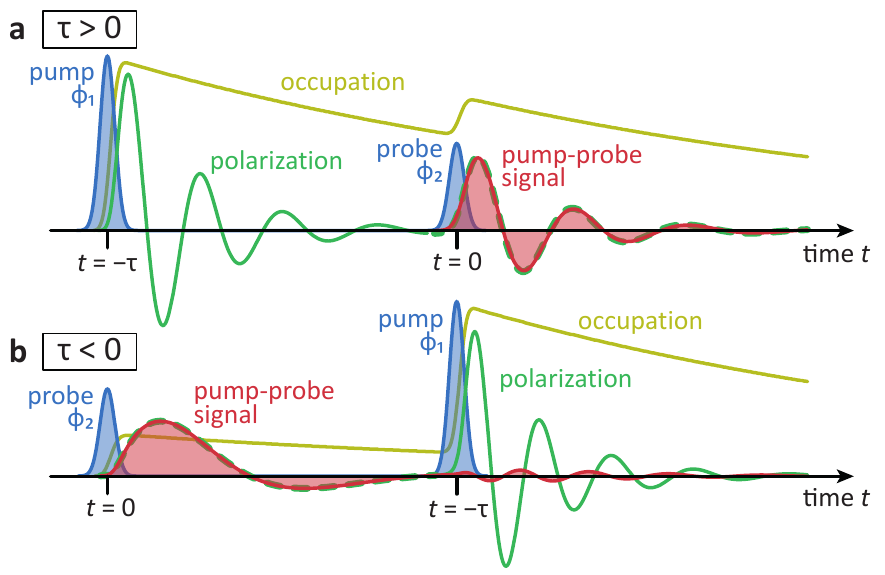}
	\caption{Schematic picture of the pump-probe signal generation. (a) For positive delays, (b) for negative delays. The signal (red) is launched by the probe pulse at $t=0$ and carries contributions from the occupation (yellow) and the polarization (green) after the pump pulse for $\tau>0$ (a). For $\tau<0$ the signal changes its dynamics by the latter impinging pump pulse (b).}
	\label{fig:scheme}
\end{figure}

In the pump-probe experiment we have to distinguish between positive and negative delays. The usual signal generated for positive delays, where the stronger pump pulse arrives before the weaker probe pulse, is schematically shown in Fig.~\ref{fig:scheme}(a). Note that the schematic depicts occupation (yellow) and polarization dynamics (green) stemming from an actual numerical simulation for a given realization of phase combination and pulse areas. It also shows a respective pump-probe signal dynamics $p_{\rm pp}$ in red. In the limit of ultrafast pulses we get in the first order of the probe pulse $\theta_2$
%%%
\begin{subequations}\begin{align}
	p_{\rm  pp}(t>0,\tau>0) &= \theta_2 e^{i\alpha(t,\tau)-\beta t} \Bigg\{ \frac{i}{2} \left[ 1-2\sin^2\left(\frac{\theta_1}2\right) e^{-\Gamma' \tau} \right]  \nonumber \\
	&+ \frac{2V - iW}{8\Gamma'}\sin^2(\theta_1) e^{-2\beta\tau - 2\kappa(\tau)}\left( 1 - e^{-\Gamma' t}\right) \Bigg\}  + \mathcal{O}(\theta_2^2) \,,
\end{align}
%%%
with
%%%
\begin{align}
	&\kappa(\tau) =  W\sin^2\left(\frac{\theta_1}2\right)\frac 1{\Gamma'} \left(1-e^{-\Gamma' \tau}\right) \,, \\
	&\alpha(t,\tau) = (V-\omega_0) t -2\frac{\widetilde{V}}{\Gamma^\prime} \sin^2\left(\frac{\theta_1}{2}\right) e^{-\Gamma^\prime \tau} \left(1-e^{-\Gamma^\prime t}\right)
\end{align}\end{subequations}
%%%
and $\widetilde{V}=V-iW/2$, which combines local field and EID in one complex quantity. The pump-probe polarization $p_{\rm pp}$ consists of two parts: The first term in the curly brackets stems from the linear polarization generated by the second pulse which automatically carries the phase $\phi_2$ and also appears in the pure 2LS without local field and EID. The second term appears because of a phase mixing of the polarization generated by the pump pulse $\sim e^{i\phi_1}$ with the phase-difference of pump and probe $\sim e^{i(\phi_2-\phi_1)}$ due to local field and EID as explained in Ref.~\cite{HahnNJP21} corresponding here to the diffraction by the transient grating generated due to the different propagation directions of the two pulses. Because both admixtures in this latter term are damped by dephasing, this contribution to $p_{\rm pp}$ decays with twice the dephasing $2\beta\tau$ and with the EID in $\kappa$ regarding the delay dynamics.

%%%%
We note that the Fourier transform of the probe pulse in Eq.~\eqref{eq:pp_signal} is a constant that is proportional to the pulse area $\theta_2$ in the delta pulse limit. With this the pump-probe signal is given by
%%%
\begin{align}
	&S(\omega,\tau>0) = \frac{1}{2}\left[ 1-2\sin^2\left(\frac{\theta_1}{2}\right) e^{-\Gamma' \tau}\right] {\rm Re} \bigg\{\mathcal{F} \left[ e^{i\alpha - \beta t} \right]\bigg\} \nonumber\\
		& + \frac{1}{8} \sin^2(\theta_1) e^{-2\kappa - 2\beta\tau} \Bigg(2V{\rm Im}\Bigg\{ \mathcal{F} \left[ \frac{1}{\Gamma'} \left(1-e^{-\Gamma' t}\right) e^{i\alpha - \beta t} \right] \Bigg\}\nonumber \\
			& \qquad\qquad - W{\rm Re} \Bigg\{  \mathcal{F} \left[ \frac{1}{\Gamma'} \left(1-e^{-\Gamma' t}\right) e^{i\alpha - \beta t} \right]  \Bigg\}\Bigg)\,.
\end{align}
%%%
This can be evaluated analytically in the case that the decay is slow in comparison to the investigated timescale, i.e.,
%%%
\begin{eqnarray*}
	\Gamma' t \ll 1
	\Rightarrow \frac{1}{\Gamma'} \left(1-e^{-\Gamma' t}\right) \approx t \,.
\end{eqnarray*}
%%%
Then the pump-probe spectrum reads
%%%
\begin{subequations}\label{eq:S_pos_tau}\begin{align}
	S(\omega,\tau>0) &= 
	 \frac{1}{2}\left[ 1-2\sin^2\left(\frac{\theta_1}{2}\right) e^{-\Gamma' \tau}\right]\frac{\tilde \beta}{\tilde \beta^2 + (\omega - \tilde\omega_0)^2} \nonumber\\
	&	 + \frac{\sin^2(\theta_1)}8 e^{-2\kappa - 2\beta\tau}\Bigg\{
		4V\frac{\tilde\beta(\omega - \tilde\omega_0)}{[ \tilde\beta^2 + (\omega - \tilde\omega_0)^2]^2} 
	+W\frac{\tilde\beta^2 - (\omega - \tilde\omega_0)^2}{[ \tilde\beta^2 + (\omega - \tilde\omega_0)^2]^2} 
		\Bigg\}\,,
\end{align}
%%%
with the effective dephasing $\tilde\beta$ and transition frequency $\tilde\omega$ resulting from the EID and local field respectively, defined as
%%%
\begin{align}
	\tilde \beta(\tau) =& \beta + W \sin^2\left(\frac{\theta_1}2\right)e^{-\Gamma'\tau}\,,\\
	\tilde\omega_0(\tau) =& \omega_0 - V + 2V\sin^2\left(\frac{\theta_1}2\right) e^{-\Gamma'\tau}\,.
\end{align}\end{subequations}
%%%
The signal consists of three contributions. In the pure 2LS ($V=W=0$, $\tilde\beta=\beta$, $\tilde\omega_0=\omega_0$) only the first term would remain, while the other two stem from EID $\sim W$ and local field $\sim V$, respectively. The first term has a Lorentzian lineshape of width $\tilde\beta$ and is centered around $\tilde\omega_0$. The EID contribution (third term) exhibits a minimum directly at and two symmetric maxima around $\omega = \tilde\omega_0$. The local field contribution (second term) is asymmetric and changes its sign from negative at $\omega < \tilde\omega_0$ to positive at $\omega > \tilde\omega_0$. Such dispersive line shapes are known to appear in few-level systems that exhibit additional internal dynamics~\cite{HenzlerPRL21}. We find that all contributions from EID and local field vanish quickly for $\tau > 0$. The overall amplitude is damped with $2\beta$ and $W$ via $\kappa(\tau)$, while $\tilde\beta$ and $\tilde\omega_0$ are additionally damped with the decay rate $\Gamma'$. Therefore, the spectral width and the spectral maximum relax exponentially towards $\beta$ and $\omega_0-V$, respectively.

At exact pulse overlap for $\tau=0^+$ the spectrum simplifies to
%%%
\begin{align}\label{eq:S_tau0}
	S(\omega,\tau=0^+) = \frac{1}{2}  \frac{\cos(\theta_1) \bar\beta }{\bar \beta^2 + (\omega - \bar\omega_0)^2} 
	 + \frac{\sin^2(\theta_1)}8 \left\{
		4V\frac{\bar\beta(\omega - \bar\omega_0)}{[ \bar\beta^2 + (\omega - \bar\omega_0)^2]^2}
		+W\frac{\bar\beta^2 - (\omega - \bar\omega_0)^2}{[ \bar\beta^2 + (\omega - \bar\omega_0)^2]^2} \right\}
\end{align}
with
\begin{align}
	\bar\beta =& \beta + W \sin^2\left(\frac{\theta_1}2\right) \,,\nonumber\\
	\bar\omega_0 =& \omega_0-\cos(\theta_1)V \,.\nonumber
\end{align}
%%%
This result already shows that for increasing pump pulse areas $\theta_1$ the dispersive, asymmetric line shape becomes more pronounced while the symmetric Lorentzian gets weaker. At the same time the spectrum gets broader and its center experiences a blue shift. Such energy shifts are well known from quantum dots~\cite{SotierNatPhys09,HunekePRB11,HinzPRB18}. All these effects are at least of second order in the pump pulse area $\theta_1$. As will be discussed in detail below, $V$ and $W$ cannot be determined independent from $\theta_1$ as long as the lowest order of the optical fields dominates. Our experiments are performed in this regime and therefore we will use $\theta_1^2V$ and $\theta_1^2W$ as fitting parameters to reproduce the experimental results. A more details discussion of possible values for $V$ and $W$ is given in Ref.~\cite{HahnNJP21}, where a connection to the parameters retrieved from microscopically derived models~\cite{Katsch2DMater19, KatschPRL20} is given.

Next we discuss the case of negative delays. Here, the probe pulse arrives first. The polarization is then simply given by the free decay of the polarization
%%%
\begin{eqnarray}\label{eq:p_tau_neg_free}
	p_{\rm pp}(0<t<-\tau,\tau<0) =  \frac i2 \theta_2 e^{-\beta t - i(\omega_0 -V)t} + \mathcal{O}(\theta_2^2) \,.
\end{eqnarray}
%%%
If the delay is large enough that the entire polarization is decayed, the spectrum is simply given by a Lorentzian of width $\beta$ centered around $\omega_0-V$. However, as schematically shown in Fig.~\ref{fig:scheme}(b) for smaller negative delays the dynamics are interrupted by the arrival of the pump pulse. Before the pump the signal dynamics follow the polarization created by the probe pulse. The appearing contributions from local field and EID change the pump-probe polarization dynamics drastically and it reads
%%%
\begin{align}
	p_{\rm pp}(t>-\tau,\tau<0) = e^{i\alpha(t+\tau, 0) - \beta t}\Bigg\{ \frac i2 \cos^2\left(\frac {\theta_1}2\right) 
	 + \frac 18 \sin^2(\theta_1) (2V - iW) \frac{1}{\Gamma'} \left[ 1-e^{-\Gamma'(t+\tau)}\right] \Bigg\} \,.
\end{align}
%%%
All in all, the polarization (i) is instantaneously reduced because the pump pulse redistributes the excitonic wave function, (ii) changes its oscillation frequency due to the local field contribution in $\alpha$ (clearly seen in Fig.~\ref{fig:scheme}(b)), and (iii) dephases faster due to the EID $W$ entering in $\alpha$. This rapid change of the temporal evolution results in spectral oscillations similar to those well known from quantum wells~\cite{FluegelPRL87,SokoloffPRB88,BorriSST03} and quantum dots~\cite{GuentherPRL02,SotierNatPhys09,HunekePRB11}, that were also reported in multi-wave coherent control experiments on individual excitons~\cite{FrasNatPhot16}. Note, that for a pure 2LS only effect (i) is present. However, by including the local field effect not only the amplitude of the signal is rapidly interrupted by the pump pulse but also the oscillation frequency and damping of the signal changes which leads to a more involved origin of the spectral oscillations.

%%%%%
%%%%%
\section{Results and Discussion}
\begin{figure*}[h!]
	\centering
	\includegraphics[width=0.75\textwidth]{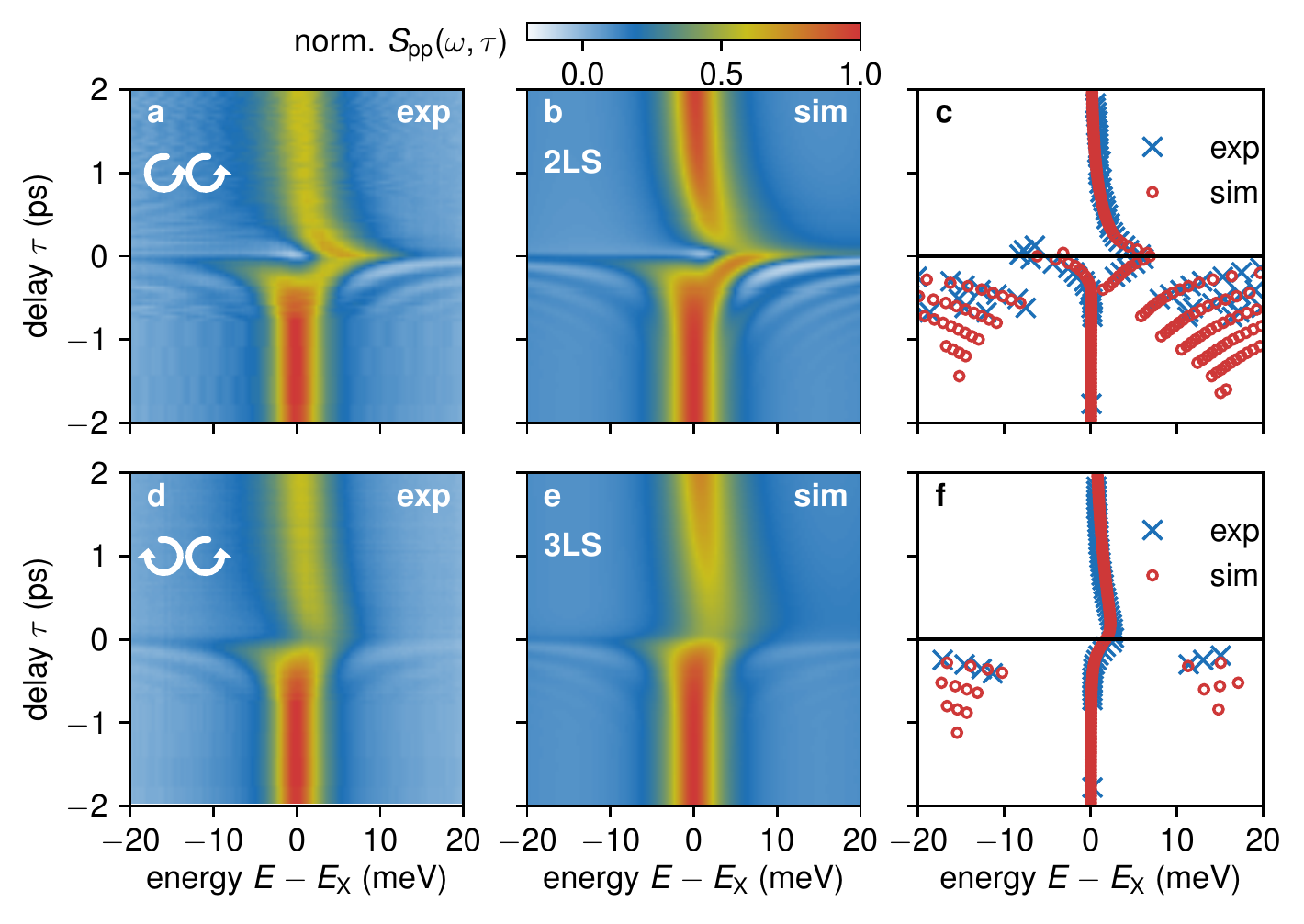}
	\caption{Delay scan of pump-probe spectra for $P = 450$~\textmu W. (a, b, c) Co-circular polarization. (d, e, f) Cross-circular polarization. (a, d) Experiment. (b, e) simulation, (b) in the 2LS and (e) in the 3LS. (c, f) Local spectral maxima in experiment (blue crosses) and simulation (red circles).}
	\label{fig:tau_scan}
\end{figure*}

%%%%%
We first perform the co-circularly polarized ($\circlearrowleft\circlearrowleft$) pump-probe experiment and vary the delay between the two pulses from $\tau=-2$~ps to 2~ps. The retrieved measured spectra are plotted in Fig.~\ref{fig:tau_scan}(a) as a function of the delay $\tau$. The corresponding simulation in the 2LS model is plotted in the same way in Fig.~\ref{fig:tau_scan}(b), where we considered a pulse duration of $\Delta t =21$~fs in agreement with autocorrelation measurements shown in the Supporting Information. In addition we choose $\hbar\theta_1^2V=\hbar\theta_1^2W=11.7$~meV to reach the overall excellent agreement with the measurement. The other parameters are fitted as independently as possible to the experiments as explained in more detail in the following. Note, that the simulations have to be performed numerically when considering non-vanishing pulse durations. Nevertheless, we will in the following use the derived equations in the ultrafast pulse limit to qualitatively explain the found spectral dynamics.

Before discussing all details of the signal's dynamics we directly see that the simulation almost perfectly reproduces the measured data. Therefore, we can analyze both plots simultaneously. As already explained in the Theory section the signal behaves entirely different for positive delays, where the probe pulse comes after the pump, and negative delays, where the signal is created already after the first arriving pulse. Starting from large negative delays $\tau\approx -2$~ps the spectrum is given by a single peak. As can be seen in Eq.~\eqref{eq:p_tau_neg_free} the width of this peak is given by the dephasing rate and is determined to $\beta=4$~ps$^{-1}$. It is therefore fitted to the experiment independently from the other parameters. Although the fitted Lorentzian agrees well with the measured spectrum, small contributions to the line width stemming from sample inhomogeneity are compensated by the choice of $\beta$. We define the center of the peak as natural exciton transition energy $E_{\rm X}$, which corresponds to $E_{\rm X}=\hbar(\omega_0-V)$ in the model. For $\tau<0$ and approaching zero the spectrum begins to develop characteristic spectral oscillations. Such features are well known for pump-probe spectra on quantum wells~\cite{FluegelPRL87,SokoloffPRB88,BorriSST03} or quantum dots~\cite{GuentherPRL02,SotierNatPhys09,HunekePRB11}, where they originate from a sudden decrease of the polarization, an increased EID by the pump pulse (arriving second) or a Coulomb-induced spectral shift, like in our case. This introduces an asymmetry in the spectrum that is observed in the shape of the oscillations. Having a close look at the spectra for $\tau\lesssim 0$ we indeed find that it splits into two peaks, which is not expected for a normal 2LS without local field effect. Focusing now on positive delays, starting at $\tau=0$ the spectrum again consists of a single pronounced maximum which is shifted to larger energies compared to $E_{\rm X}$. For increasing delays this maximum moves back to its original energy within approximately 2~ps. At the same time we see that the intensity starts with a maximum and is slightly quenched for a short time interval during this relaxation process, both in experiment and theory.

%%%%
%%%%

The entire spectral dynamics are quite involved, especially for negative delays where spectral oscillations build up. Although the two color plots look alike we try to find a more quantitative comparison between experiment and theory for the entire delay scan. For that purpose in Fig.~\ref{fig:tau_scan}(c) we plot the positions of local maxima in the spectra. The found positions are plotted as blue crosses for the measurement and as red circles for the simulation. We find that the traces of local spectral maxima match perfectly, which demonstrates the high accuracy of our model. Considering the spectral relaxation for $\tau>0$ we again have a look at the analytical Eqs.~\eqref{eq:S_pos_tau} for ultrafast pulses and find that the signal shift decreases exponentially with the effective decay rate $\Gamma'$. Therefore, we use this decay to determine the rate to $\Gamma'=1.6$~ps$^{-1}$ again independently from the other system parameters. More details on this population relaxation are given in the Supporting Information. The found strong similarity between the co-circular measurement and the calculation in the 2LS show that it is a justified approximation for this excitation scheme. In the Supporting Information we also directly compare the numerical and analytical simulations in the 2LS.

Next, we move to the cross-circularly polarized ($\circlearrowleft\circlearrowright$) pump-probe experiment, which is depicted in Fig.~\ref{fig:tau_scan}(d) in the same way as the co-polarized one. The fact that we detect spectral oscillations and a spectral shift of the signal maximum around $\tau=0$ shows that we have to consider the entire 3LS in Fig.~\ref{fig:levels}(a). A cross-polarized pulse sequence cannot be described in a 2LS model. The corresponding simulation is depicted in Fig.~\ref{fig:tau_scan}(e) and the direct comparison of the local spectral maxima is given in Fig.~\ref{fig:tau_scan}(f). To achieve, once more, a remarkable agreement between simulation and measurement we determine the exciton decay to $\Gamma= 0.6$~ps$^{-1}$, the dephasing to $\beta=4$~ps$^{-1}$ (same as for $\circlearrowleft\circlearrowleft$) and adjusted the pulse area slightly such that the parameters are now given by $\hbar\theta_1^2V=\hbar\theta_1^2W=13$~meV. The reason for the changed pulse area is at least partly due to a renormalization of the pulse area when changing the decay channels between 2LS and 3LS as discussed in more detail in the Supporting Information. To find the best agreement, we set the inter-valley scattering rate to $\lambda=4.4$~ps$^{-1}$. As expected, we find that the effective decay rate in the 2LS is larger than the one in the 3LS. While in principle the same features as in the co-polarized case are found, they are significantly less pronounced. When the occupation is transferred between the valleys, the maximal occupation in the not-pumped valley does not exceed half of the maximal occupation in the pumped valley. Therefore the energy shifts are expected to be much smaller in cross-polarized excitations. One remarkable qualitative difference is found for small positive delays around the maximal energy shift of the spectral maximum, where the cross-polarized spectrum is strongly suppressed. The reason is that the phase-mixing processes between pump and probe, that enter in Eq.~\eqref{eq:S_pos_tau} via the combination of $\phi_2-\phi_1$ and $\phi_1$, are not possible because the two pulses are orthogonally polarized. Therefore, the probe polarization is simply experiencing an additional oscillation due to the local field from the occupation already transferred into the probed valley. The signal is significantly damped due to the dephasing and the EID which depends on the total occupation. This total occupation is preserved under inter-valley scattering and decays only with the rate $\Gamma$ and the damping influence of the EID is therefore maximal for $\tau=0$. This finally results in the significant suppression of the signal at small $\tau$.

Although a cross-polarized excitation in principle allows for a biexciton creation, we do not find pronounced spectral features in the measured pump-probe spectra characteristic for a biexcitonic transition. On the one hand, the reason might be that the biexciton binding energy of several tens of meV~\cite{HaoNatComm17, KezerashviliFBS17} is large enough, such that the respective optical transition is not efficiently driven by the laser pulse centered at the exciton line. On the other hand, calculations for WSe$_2$ within a microscopic model show that the biexciton contribution is strongly broadened and mainly consists of a weak shoulder on the low-energy side of the exciton line~\cite{KatschPRL20}.

To confirm the consistency of our model we performed the simulations for the co-circularly polarized excitation also in the 3LS model considering the same system parameters as for the cross-circular excitation. By slightly adjusting the pump pulse areas to $\hbar\theta_1^2V=\hbar\theta_1^2W=13$~meV we achieved an equally good agreement with the measurement as in the 2LS model depicted in Figs.~\ref{fig:tau_scan}(a) -- (c). The corresponding simulations in the 3LS for co-polarized excitation are shown in the Supporting Information.

\begin{figure*}[h!]
	\centering
	\includegraphics[width=0.75\textwidth]{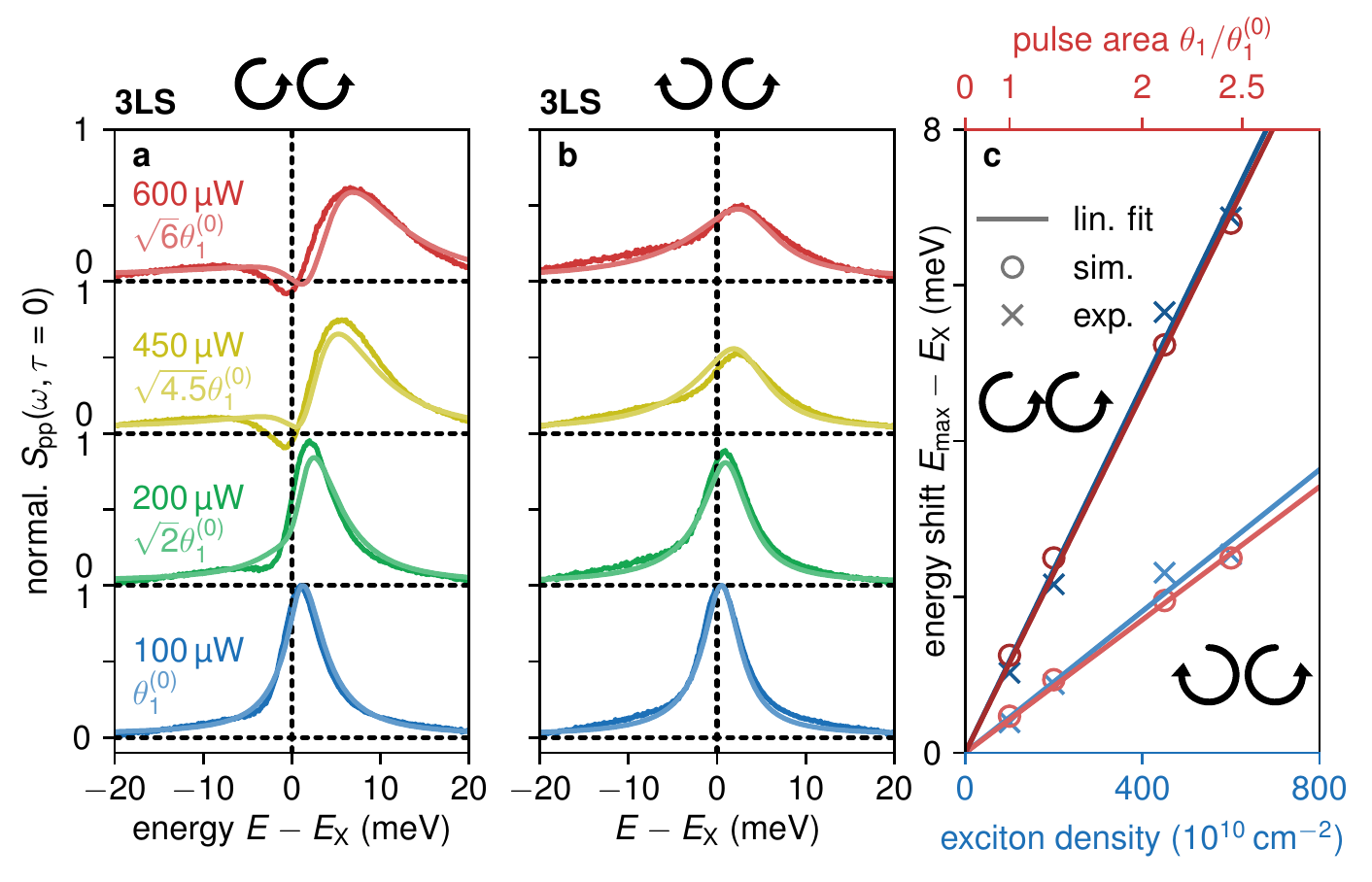}
	\caption{Pump-probe spectra at pulse overlap. (a) Co-circular polarization with experiment in dark and simulation in the 3LS in bright colors. For the smallest pulse area we have $\hbar{\theta_1^{(0)}}^2 V = \hbar{\theta_1^{(0)}}^2W = 2.4$~meV. (b) Cross-circular polarization with simulations in the 3LS, with $\hbar{\theta_1^{(0)}}^2 V = \hbar{\theta_1^{(0)}}^2W = 3.2$~meV. (c) Positions of the spectral maxima against the exciton density (bottom axis) in the experiment and the pulse area (top axis) in the simulation. An exciton density of $10^{12}$/cm$^2$ corresponds to an average pumping power of about 100~\textmu W (see Supporting Information).}
	\label{fig:tau0}
\end{figure*}

%%%%%%%%%
As shown in Eq.~\eqref{eq:S_tau0} we expect that the influence of local field and EID manifest in the particular spectral shape at $\tau=0$ in the 2LS. In addition we find that their influence can be controlled by changing the pump pulse area $\theta_1$. In Fig.~\ref{fig:tau0}(a), we present typical pump-probe spectra with co-circular polarization ($\circlearrowleft\circlearrowleft$) obtained at pulse overlap, i.e., a delay of $\tau=0$, for increasing intensities of the pump pulse from bottom to top. The energy axis is shifted by $E_{\rm X}$, i.e., to the peak energy at the largest negative delay of $\tau=-2$~ps and all depicted curves are normalized in amplitude to the first spectrum (blue). The dark lines show the measured data and the light ones the simulations based on the 3LS (a corresponding simulation in the 2LS is shown in the Supporting Information). Starting with the smallest considered excitation power we find a single nearly symmetric peak that is slightly shifted to energies larger than $E_{\rm X}$. With increasing pulse powers we find four striking changes of the spectra: (i) The peak maximum moves to larger energies, reaching a shift of approximately 6.5~meV for a pump power of 600~\textmu W, (ii) the peak intensity shrinks by nearly 40\% for the largest power, (iii) the peak gets significantly broader, and (iv) the spectrum develops an increasingly pronounced dispersive feature that even reaches negative values for the two largest pulse powers.

%%%%
%%%%
 
Comparing the measured spectra with the simulations we obviously find a strong similarity for all considered pump powers. The fitted system parameters are $\Gamma=0.6$~ps$^{-1}$, $\lambda=4.4$~ps$^{-1}$ (both the same as in Fig.~\ref{fig:tau_scan}(d)-(f)), and $\beta = 3$~ps$^{-1}$. In the Supporting Information we directly compare Fig.~\ref{fig:tau0}(a) with Eq.~\eqref{eq:S_tau0} to additionally confirm that our analytic calculations in the delta-pulse limit accurately describe the numerical simulations with a pulse duration of $\Delta t=21$~fs. There, we find only slight deviations in the local field induced energy shift between the analytic and the numeric results. The considered pulse areas in the simulation are listed in Fig.~\ref{fig:tau0}(a) next to each spectrum. Starting from the lowest pulse area $\theta_1^{(0)}$ corresponding to 100\,\textmu W the higher areas grow according to the increase in pulse power in the experiment. As the pulse area is proportional to the electric field $E$, the pulse power scales with its intensity $E^2$.

To extract a quantitative measure from the set of spectra we consider finding (i) and determine the energy $E_{\rm max}$ of the pronounced maximum in the spectrum by fitting a Lorentzian locally around the maximum and plot this quantity as a function of excitation density in Fig.~\ref{fig:tau0}(c). The blue crosses give the experimental and the red circles the theoretical data. As expected from the prediction in Eq.~\eqref{eq:S_tau0} we find a linear shift of the spectral resonance for small pulse intensities $\sim\theta_1^2$. At this point it is important to note again that the influence of local field and EID enter the model in the lowest order of the pump field via $\theta_1^2V$ and $\theta_1^2 W$, respectively. Therefore, when operating in this regime of the optical fields, shown by the linear fit in Fig.~\ref{fig:tau0}(c), it is not possible to determine the actual pulse area and $V$, $W$ independently. This means that in the simulations the choice of a larger $\theta_1$ can be compensated by smaller $V$ and $W$. However, we can determine the product of pulse power and local field factor, respectively EID strength, to $\hbar{\theta_1^{(0)}}^2 V = \hbar{\theta_1^{(0)}}^2W = 2.4$~meV.

Also the other findings (ii) -- (iv) can be traced back to results from our theoretical model in Eq.~\eqref{eq:S_tau0}. (ii) The fading of the signal strength stems from the term $\sim - \sin(\theta_1/2)W$ reducing the amplitude of the Lorentzian contribution of the spectrum due to the EID. (iii) The same effect leads to the increasing width of the spectrum in $\bar\beta$ [Eq.~\eqref{eq:S_tau0}] such that the EID approximately preserves the integrated intensity. Considering Eq.~\eqref{eq:S_tau0} we find that the total intensity decreases $\sim\theta_1^2$ in the lowest order, as the integral over the last two terms vanishes. (iv) The appearance of a minimum in the spectrum can directly be traced back to the local field effect contributing with a dispersive feature in Eq.~\eqref{eq:S_tau0}. These findings of the influence of local field and EID depending on the excitation density are in agreement with Refs.~\cite{Katsch2DMater19, KatschPRL20}.

In Fig.~\ref{fig:tau0}(b) we perform the same pump power analysis for the cross-circular excitation ($\circlearrowleft\circlearrowright$) where we consider $\hbar{\theta_1^{(0)}}^2 V = \hbar{\theta_1^{(0)}}^2W = 3.2$~meV and keep $\Gamma$, $\lambda$, and $\beta$ fixed. As for the full delay scan in Fig.~\ref{fig:tau_scan} we find that the energy shifts are less pronounced than in the co-polarized case. Also no clear minimum develops in the spectrum at increased excitation powers. However, the spectral line still significantly broadens, which shows that the influence of the EID remains nearly unperturbed. This demonstrates that the effect is only depending on the total exciton density as we have considered in our model in Eq.~\eqref{eq:3LS}. At the same time the local field depends on the polarization orientation of the excitons. Therefore, the induced energy shift is much less pronounced in a cross-polarized excitation scenario because the probed valley has to get occupied to develop a local field coupling. This finding is again summed up in Fig.~\ref{fig:tau0}(c) where the energy shifts for $\circlearrowleft\circlearrowright$ exhibit nearly half the slope of $\circlearrowleft\circlearrowleft$.

\section{Conclusions and Outlook}
In summary, we have studied the shape of the absorption features in the hBN/MoSe2/hBN heterostructure in the regime of ultrafast resonant excitation. All characteristic optical signatures related to exciton dynamics that appear in the pump-probe experiments were fully reproduced by the applied local-field model. By combining experiment and theory we have demonstrated that for pulse delays shorter than the exciton lifetime the A exciton response shows a blue shift of a few meV, which we could trace back to the impact of the local field effect. In addition, by investigating spectral line widths and amplitudes we could study the influence of excitation induced dephasing effects. It was found that both effects approximately have the same strength and also influence the appearance of spectral oscillations for an inverted pulse ordering. When moving from a co-circularly polarized excitation scheme to cross-circular excitation we had access to the inter-valley scattering rate which we found to be in the range of a few ps$^{-1}$. 

Our results are in line with previous nonlinear spectroscopy studies of TMD systems and further strengthen the potential of this technique to explore ultrafast dynamics in layered materials. A handy few-level model, which even allows to develop analytical expressions in some special cases, explains the observed features at least qualitatively, thus offering insight into the physics behind the spectral dynamics. Therefore, we have a useful tool for example to analyze higher excited states like biexcitons or the fundamental impact of external magnetic fields in the context of local field coupling and excitation induced dephasing. Such external perturbations lead to distinct shifts of the exciton energies in opposite valleys, which have an impact on inter- and intra-valley scattering mechanisms. In the forthcoming work, we will explore the impact of such high exciton density effects on coherent optical response in more advanced experimental and theoretical configurations, for example offered in four-wave mixing spectroscopy.

%\begin{acknowledgement}
%  Please insert acknowledgments of the assistance of colleagues or similar notes of appreciation here.
%\end{acknowledgement}

\begin{funding}
The Warsaw team has been supported by the ATOMOPTO project (TEAM programme of the Foundation for Polish Science, co-financed by the EU within the ERDFund), CNRS via IRP {\it 2D~Materials}, EU Graphene Flagship. The Polish participation in the European Magnetic Field Laboratory (EMFL) is supported by the DIR/WK/2018/07 of MEiN of Poland. A.R. acknowledge support from the {\it Diamentowy Grant} under decision  DI2017 008347 of MEiN of Poland. T.H., T.K., P.M., and D.W. acknowledge support from the Polish National Agency for Academic Exchange (NAWA) under an APM grant (No. PPI/APM/2019/1/00085). T.H. thanks the German Academic Exchange Service (DAAD) for financial support (No. 57504619). D.W. thanks NAWA for financial support within the ULAM program (No. PPN/ULM/2019/1/00064). J.K. acknowledges the support from Integrated Development Programme of Warsaw University. K.W. and T.T. acknowledge support from the Elemental Strategy Initiative conducted by the MEXT, Japan, (grant no. JPMXP0112101001), JSPS KAKENHI (grant no. JP20H00354), and the CREST (JPMJCR15F3), JST.
\end{funding}

\bibliographystyle{unsrtnat}
%\bibliography{pp.bib}

\end{document}

%%%--------------------------------------------%%%
	\articletype{Research Article \hfill Accepted version of Nanophotonics 10, 2717-2728 (2021)}
%	\received{Month	DD, YYYY}
%	\revised{Month	DD, YYYY}
%  \accepted{Month	DD, YYYY}
 % \journalname{De~Gruyter~Journal}
 % \journalyear{YYYY}
 % \journalvolume{XX}
 % \journalissue{X}
  \startpage{1}
%  \aop
 % \DOI{10.1515/sample-YYYY-XXXX}
%%%--------------------------------------------%%%

\title{Local field effects in ultrafast light-matter interaction measured by pump-probe spectroscopy of monolayer MoSe$_{\boldsymbol 2}$\\ (Supporting Information)}
\runningtitle{Exciton dynamics in MoSe$_{2}$ (SI)}
%\subtitle{(Supporting Information)}

\author*[1]{Aleksander~Rodek}
%\ use * to mark the author as the corresponding author
\author[2,3]{Thilo~Hahn}
\author[1,4]{Jacek~Kasprzak}
\author[1]{Tomasz~Kazimierczuk}
\author[1]{Karol~Nogajewski}
\author[1]{Karolina~Po\l{}czy\'nska}
\author[5]{Kenji~Watanabe}
\author[5]{Takashi~Taniguchi}
\author[2]{Tilmann~Kuhn}
\author[3]{Pawe\l{}~Machnikowski}
\author[1,6]{Marek~Potemski}
\author[3]{Daniel~Wigger}
\author[1]{Piotr~Kossacki}
 
\runningauthor{A.~Rodek, T. Hahn et al.}

\affil[1]{\protect\raggedright 
Institute of Experimental Physics, Faculty of Physics, University of Warsaw, 02-093 Warszawa, Poland, e-mail: aleksander.rodek@fuw.edu.pl}
\affil[2]{\protect\raggedright 
Institut f\"ur Festk\"orpertheorie, Universit\"at M\"unster, 48149 M\"unster, Germany, e-mail: t.hahn@wwu.de}
\affil[3]{\protect\raggedright 
Department of Theoretical Physics, Wroc{\l}aw University of Science and Technology, 50-370 Wroc{\l}aw, Poland}
\affil[4]{\protect\raggedright 
Universit\'{e} Grenoble Alpes, CNRS, Grenoble INP, Institut N\'{e}el, 38000 Grenoble, France}
\affil[5]{\protect\raggedright 
National Institute for Materials Science, Tsukuba, Ibaraki 305-0044, Japan}
\affil[6]{\protect\raggedright 
Laboratoire National des Champs Magn\'{e}tiques Intenses, CNRS-UGA-UPS-INSA-EMFL, 38042 Grenoble, France}

%\communicated{...}
%\dedication{...}
	
%\abstract{\\
\noindent\textbf{This document includes following content:\\[2mm]
1. Co-circular simulations in the three-level system\\
2. Simulations in the ultrafast pulse limit\\
3. Impact of decay channels\\
4. Estimation of carrier concentrations\\
5. Determination of the pulse duration
}
%}

%\keywords{transition metal dichalcogenide monolayer, nonlinear spectroscopy, local field effect}

\maketitle
	
%%%%%%%%%%%%%
\section{Co-circular simulations in the three-level system}
\begin{figure}[h!]
	\centering
	\includegraphics[width=\columnwidth]{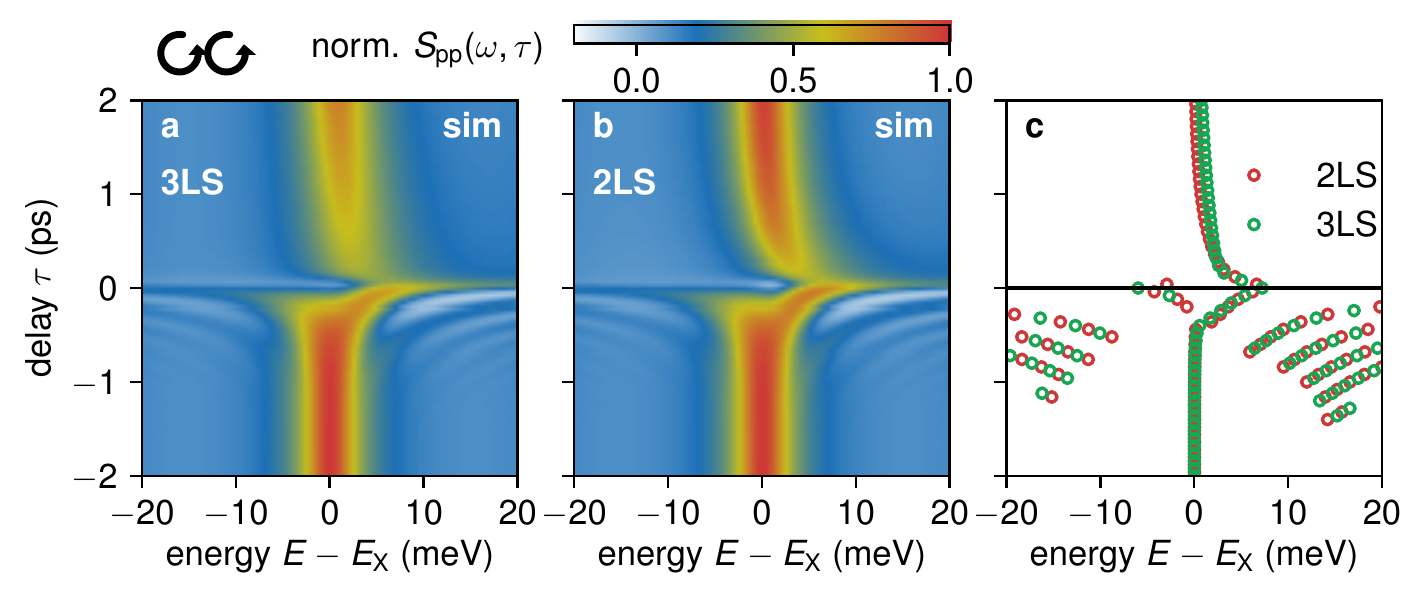}
	\caption{Delay scan of simulated pump-probe spectra for co-circular polarization. (a) Simulation in the 3LS and (b) in the 2LS. (c) Local spectral maxima in the 2LS (red circles) and the 3LS (green circles). }
	\label{fig:tau_scan_SI}
\end{figure}

Figure~\ref{fig:tau_scan_SI}(a) shows simulations for the co-circularly polarized excitation carried out in the three-level system (3LS). These are compared to cooresponding simulations in the two-level system (2LS) in Fig.~\ref{fig:tau_scan_SI}(b), which are the same as in Fig.~4(b) in the main text. Regarding the spectral oscillations for negative delays and the energy shift and relaxation for positive delays, which are highlighted in Fig.~\ref{fig:tau_scan_SI}(c), we find a good agreement. This shows that the experiment is well reproduced in both models. However, we find one slight difference in the 3LS compared to the 2LS: The amplitude of the signal recovers significantly slower for positive delays after it is quenched for $\tau\gtrsim 0$. This leads to an overall better agreement with the experiment in Fig.~4(a) in the main text. The reason for this is that the decay rate in the 3LS is smaller than the effective one in the 2LS $\Gamma < \Gamma'$. Given that the scattering does not reduce the total occupation and that the EID depends on this total exciton occupation, the influence of the EID decays much slower in the 3LS. Therefore the signal recovery takes much longer in the 3LS.
%%%%
%%%%

\begin{figure}[h!]
	\centering
	\includegraphics[width=\columnwidth]{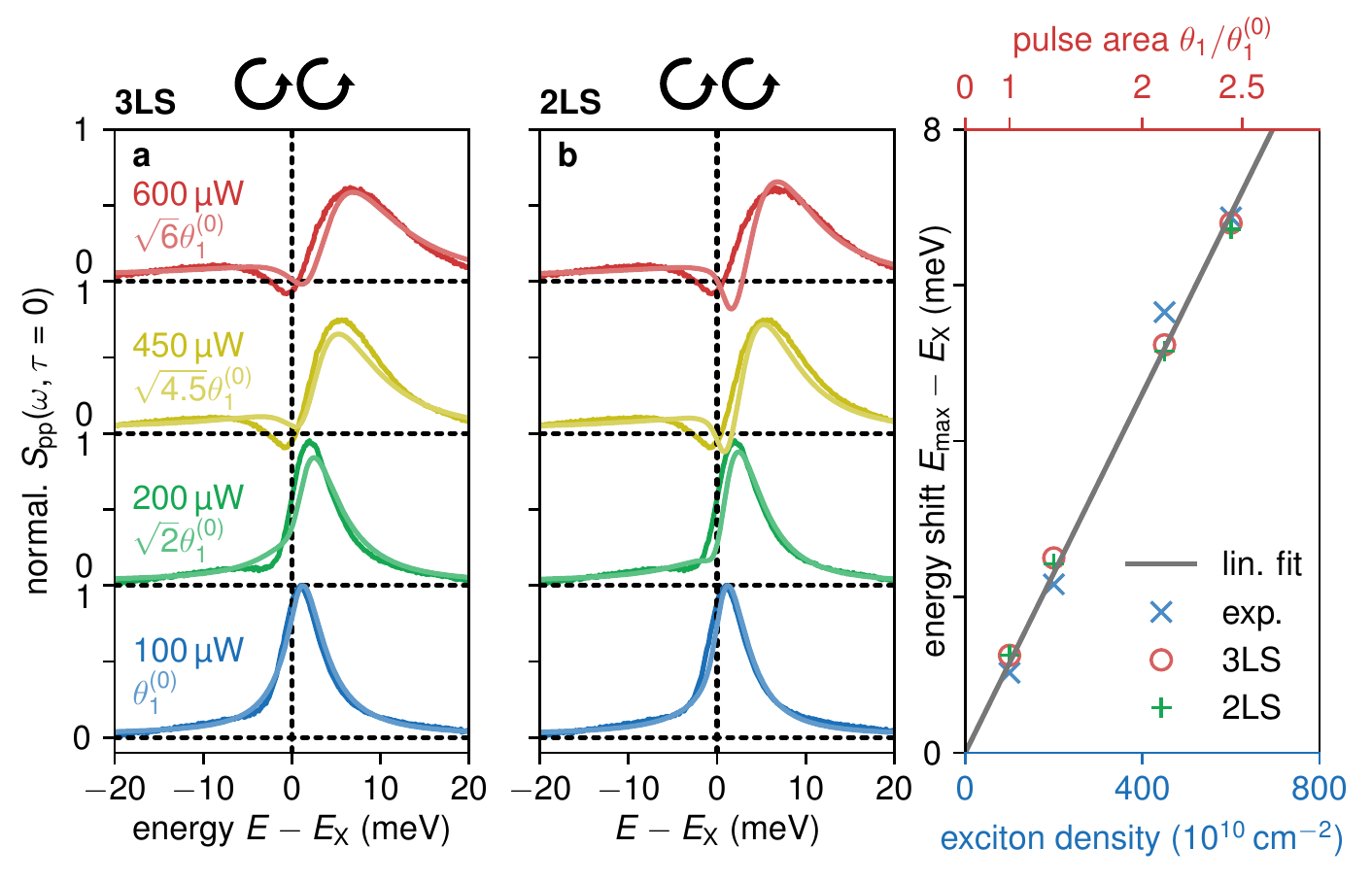}
	\caption{Pump-probe spectra at pulse overlap. (a) Co-circular polarization with experiment in dark and simulation in the 3LS in bright colors. The smallest pulse area was determined to $\hbar{\theta_1^{(0)}}^2V=\hbar{\theta_1^{(0)}}^2W=2.2$~meV. (b) Same as (a) but in the 2LS (Fig.~5(a) in the main text). (c) Positions of the spectral maxima against the exciton density (bottom axis) in the experiment and the pulse area (top axis) in the simulation.}
	\label{fig:tau0_SI}
\end{figure}

To demonstrate that also the pump-probe spectra at $\tau=0$ in co-circular polarization can be well reproduced in the 2LS Fig.~\ref{fig:tau0_SI}(a) shows the respective results. Panel (b) is the same as Fig.~5(a) in the main text. To reach the excellent agreement between simulation (bright) and measurement (dark) we determined the system parameters to $\Gamma'=1.6$~ps$^{-1}$, $\beta=3$~ps$^{-1}$, and $\hbar{\theta_1^{(0)}}^2V=\hbar{\theta_1^{(0)}}^2W=2.2$~meV. We again find that the effective decay rate has to be increased with respect to $\Gamma$ in the 3LS to compensate for the missing inter-valley scattering channel. The energy shift as a function of the pump intensity is depicted in Fig.~\ref{fig:tau0_SI}(c) and shows an excellent agreement between the 2LS, the 3LS simulations and the experiment.
%%%%
%%%%

%%%%%%%%%%%%%
\section{Simulations in the ultrafast pulse limit}
\begin{figure}[h!]
	\centering
	\includegraphics[width=\columnwidth]{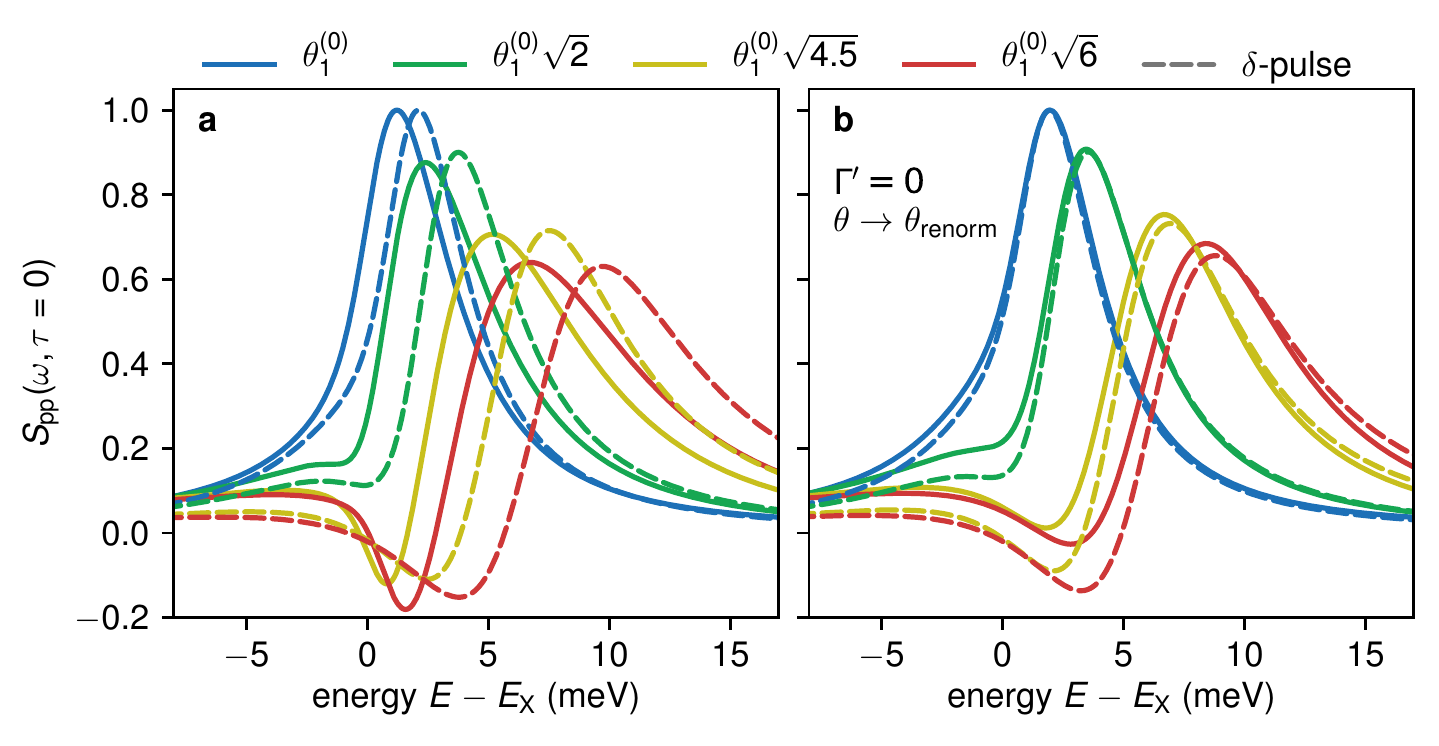}
	\caption{Comparison of simulated spectra for vanishing delay $\tau=0$. (a) Numerical simulations for a pulse duration $\Delta t=21$~fs from Fig.~5(a) in the main text in sold. Corresponding analytical results from Eq.~(11) in the main text. (b) Same as (a) but with $\Gamma'=0$ and renormalized pulse areas.}
	\label{fig:2LS_delta_tau0_SI}
\end{figure}

In Fig.~\ref{fig:2LS_delta_tau0_SI} we directly compare the numerically simulated spectra at pulse overlap, i.e., $\tau=0$, in the 2LS with the ones from the analytical derivations. The solid lines stem from the numerical simulation and are the same as in Fig.~5(a) in the main text. The dashed lines are the respective results in the ultrafast pulse limit in Eq.~(11) in the main text. In Fig.~\ref{fig:2LS_delta_tau0_SI}(a) we choose exactly the same parameters for the numerical and the analytical calculations. Overall we find that the spectral shapes agree very well for each considered pump pulse area. But we find that each spectrum in the delta-pulse limit shows a larger shift to higher energies. There are two reasons for this: (i) In the analytic result we entirely disregard the exciton decay ($\Gamma'=0$), which leads to a larger occupation and therefore a larger local field induced shift. (ii) During the excitation in the numerical calculation the system already dephases, which results in reduced polarization and occupation compared to the analytic result. This additionally results in an increased energy shift as explained before. To confirm this in the numerical simulations in Fig.~\ref{fig:2LS_delta_tau0_SI}(b) be set $\Gamma'=0$ and choose a renormalized pulse area to compensate for the dephasing during the pulse  (see following section for more details). Now we indeed find an almost perfect agreement between the two sets of simulations. The remaining deviations stem from the still non-vanishing pulse durations of $\Delta t=21$~fs in the numerical treatment. If this value is reduced to approximately 1~fs both simulations agree perfectly (not shown here).
%%%%
%%%%

%\newpage

%%%%%%%%%%%%%
\section{Impact of decay channels}

\begin{figure}[h!]
	\centering
	\includegraphics[width=\columnwidth]{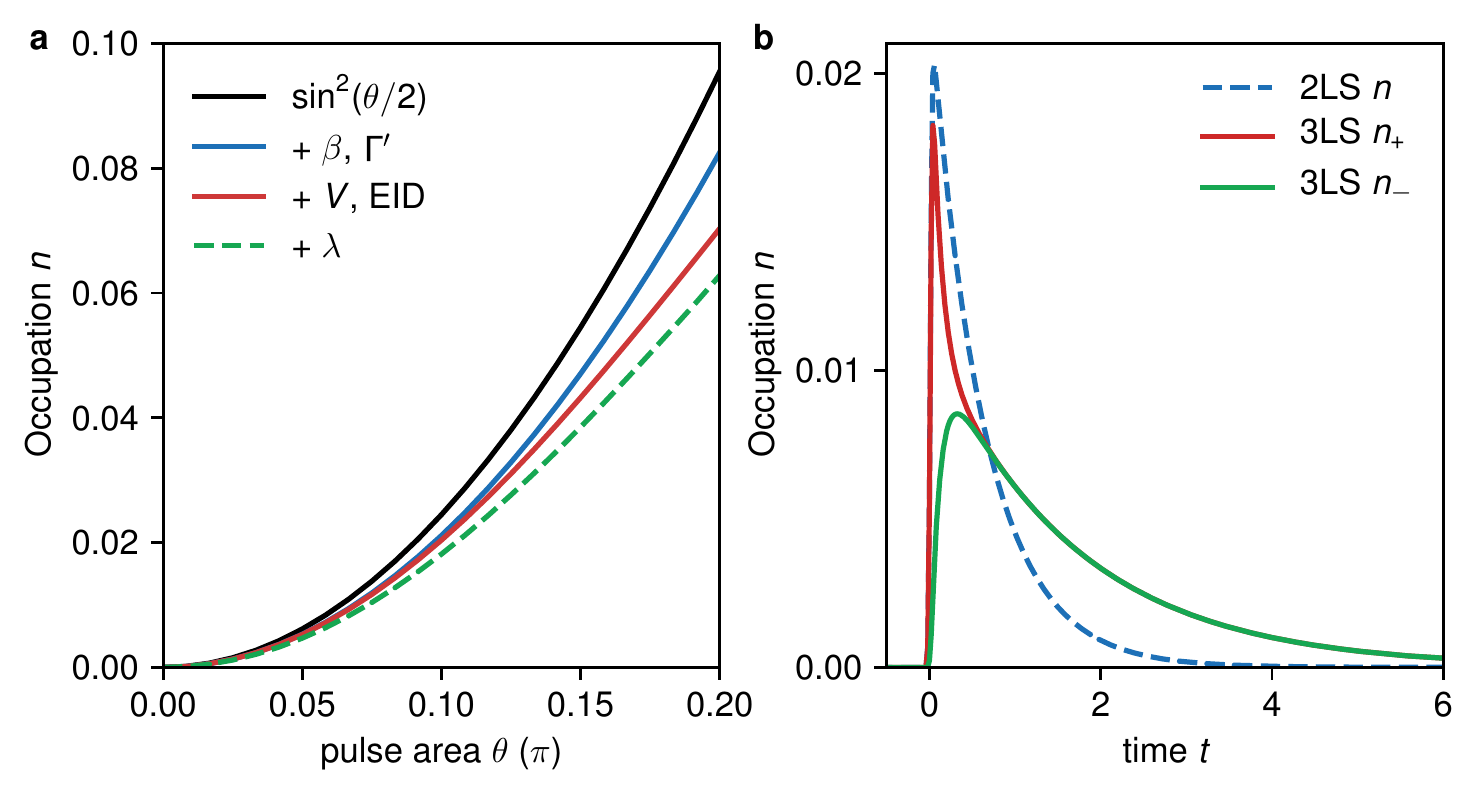}
	\caption{(a) Pulse area dependence of the exciton occupation in the 2LS and 3LS model considering different contributions in the system dynamics. Black: pure 2LS in the ultrafast pulse limit, blue: including a non-vanishing pulse duration and dephasing rate $\beta$ and effective decay rate $\Gamma'$, red: additionally including local field and EID, dashed green: adding inter-valley scattering int he 3LS. (b) Occupation decay after a pulse excitation in the 2LS (dashed blue) and the 3LS in red for $n_+$ and green for $n_-$.}
	\label{fig:pulse_areas_SI}
\end{figure}

As briefly discussed in the Theory section of the main text, the occupation after a single pulse depends on the considered system. As a reference we choose the pure 2LS without any decay excited with an ultrafast pulse where the occupation is given by $n=\sin^2(\theta/2)$, which is shown as black solid line in Fig.~\ref{fig:pulse_areas_SI}(a). When considering a non-vanishing pulse duration of $\Delta t=21$~fs and adding the dephasing rate $\beta$ and the effective decay rate $\Gamma'$ from Fig.~5(a) in the main text we have additional contributions reducing the occupation after the pulse. This leads to the solid blue curve which clearly exhibits smaller occupations for a given pulse area. The occupation is further reduced when local field $V$ and EID $W$ from Fig.~5(a) are included in the red curve. Finally going to the 3LS by including the inter-valley scattering $\lambda$, and choose $\Gamma$ from Fig.~5(b) (main text) instead of $\Gamma'$ we get the green dashed curve with the smallest occupations. This discussion explains why we needed to choose smaller pulse areas in the 2LS than in the 3LS because be needed to reach approximately the same occupation in the different systems. It also shows why the simulations in the ultrafast pulse limit in Fig.~\ref{fig:2LS_delta_tau0_SI} exhibit a larger energy shift because we chose the same pulse area as for the non-vanishing pulse duration.
%%%%
\\
%%%%
~

To again embark on the interplay between the inter-valley scattering $\lambda$ and the decay rates $\Gamma^{(\prime)}$ in Fig.~\ref{fig:pulse_areas_SI}(b) we plot the occupation dynamics after a pulse excitation at $t=0$. The dashed blue curve depicts the 2LS with effective $\Gamma'$, while the solid red and green curve show $n_+$ and $n_-$ in the 3LS, respectively. The laser pulse only addresses the $\left|+\right>$ exciton in the 3LS, which after the excitation rapidly scatters into the $\left|-\right>$ exciton. After the two occupations are balanced the decay slows down significantly. As explained in the Theory section of the main text the loss rate changes over time due to the scattering process. To compensate for this time dependence in the 2LS we choose the effective decay rate $\Gamma'$ such that the mono-exponential decay is slower at the beginning and faster at the end of the relaxation as seen in Fig.~\ref{fig:pulse_areas_SI}(b).

%%%%%%%%%%%%%
\section{Estimation of carrier concentrations}
\begin{figure}[h!]
	\centering
	\includegraphics[width=\columnwidth]{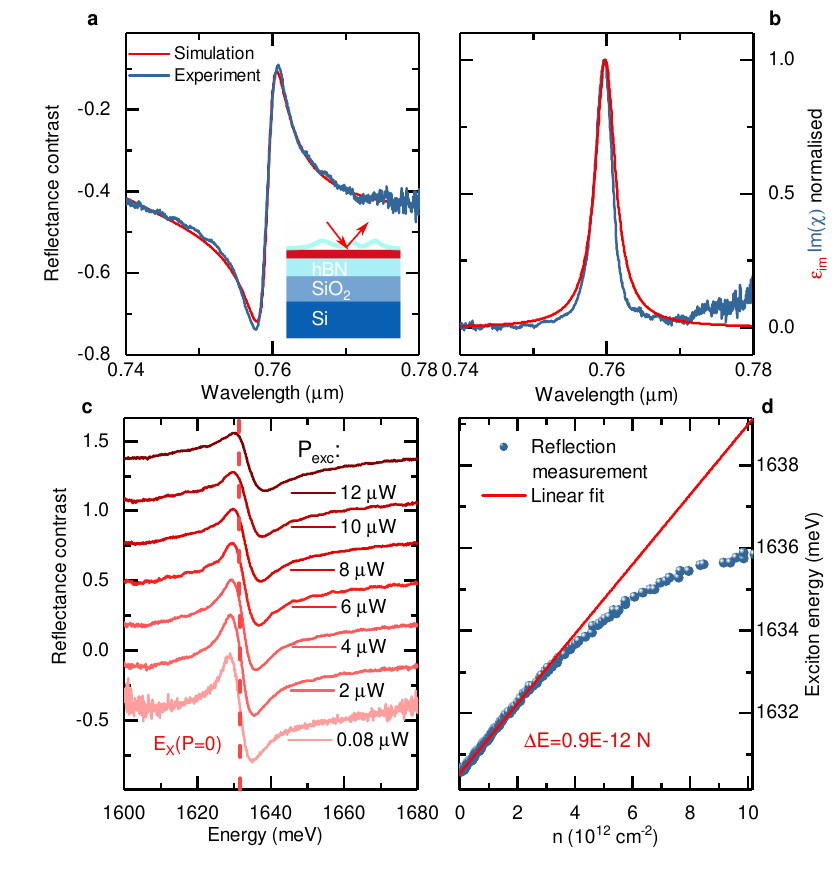}
	\caption{(a) Reflectance contrast of the MoSe$_2$ monolayer and transfer-matrix method simulation. Simulation parameters: $d_{{\rm SiO}_2}=90$~nm, $d^{\rm hBN}_{\rm bot}=85$~nm, $d^{\rm hBN}_{\rm top}= 30$~nm, $d_{{\rm MoSe}_2}= 0.7$~nm, $d_{\rm air}= 70$~nm, $n_{\rm hBN}= 2.1$, $n_{{\rm SiO}_2}= 1.54$, and $n_{\rm Si}= 3.9$. Inset: Structure of the sample. (b) Normalized imaginary part of permittivity fitted from the TMM simulation and imaginary susceptibility extracted from the spectrum by the Kramers-Kronig transformation as mentioned in the main text. (c) Reflectance contrast of the sample measured for different powers of the fs laser. (d) Energy of neutral exciton resonance as a function of the density of photogenerated carriers.}
	\label{fig:concentration}
\end{figure}

One of the consequences of the spatial separation of pump probe laser beams in the setup is the misalignment of the pump beam with the optical axis of the lens. This in turn results in a larger excitation spot on the sample, which on the one hand fully encompasses the probing area, but on the other hand makes it difficult to estimate the actual density of photogenerated excitons. In order to calibrate the concentration of carriers during the pump-probe measurement we measured the reflection from the sample with a single probing beam as a function of its power. The path of the probe laser beam lies on the optical axis of the lens allowing for good control over the size and shape of the laser spot on the surface of the sample. The final results presented in Fig.~\ref{fig:concentration}(c) show a pronounced blueshift of the exciton resonance that increases with the power of the laser. From this we can extract how it depends on the concentration. To do so  we calculate the photogenerated carriers via $n=\gamma n_{\rm photon}/(\pi r^2)$, where $\gamma$ is the total absorption of the ML, $n_{\rm photon}=PT/E_{\rm avg}$ is the number of photons per pulse with the laser power $P$, the laser repetition period $T$, the average energy of photons $E_{\rm avg}$, and the laser spot radius $r$.

To calculate the absorption one needs to look at the overlap of the ML absorption coefficient and the fs pulse spectrum. In order to take into account contributions originating from light interfering between different layers of our sample we simulate the reflection by a transfer matrix method (TMM) similarly to previous works \cite{ScuriPRL18, BackPRL18}. In this approach the total reflection from the heterostructure is given by the ratio of transfer matrix elements $R=|M_{21}/M_{11}|^2$. Here, $M$ is the product of all successive interface and layer matrices of the full heterostructure. The propagation of light on a single interface is given by 
\begin{equation}
M_{\rm interface}(n_1,n_2)=\frac{1}{2n_2}\begin{pmatrix}
n_1+n_2 & -n_1+n_2\\
-n_1+n_2 & n_1+n_2
\end{pmatrix}
\end{equation}
where $n_1$ and $n_2$ are the refraction coefficients of neighboring materials 1 and 2. The propagation within a layer is described by
\begin{equation}
M_{\rm layer}=\begin{pmatrix}
e^{ikn_id_i} & 1\\
1 & e^{-ikn_id_i} 
\end{pmatrix}
\end{equation}
where $k$ is the wavevector in vacuum and $n_i$ the refraction index and $d_i$ the thickness of layer $i$.

In the simulation we use the following values for the refractive indexes of the heterostructure materials: n$_{\rm hBN}$= 2.1, n$_{\rm SiO_2}$= 1.54, n$_{\rm Si}$= 3.9 \cite{LeePSSB19, MalitsonJOSA65, AspnesPRB83}.
In Fig.~\ref{fig:concentration}(a) we present the measured reflectance contrast from the heterostructure (blue) and the TMM simulation (red). The fitting was done via the resonance parameters included as an imaginary addition to the permittivity function of an isolated optical transition
$$\epsilon_{\rm Im} = k \frac{g}{(E-E_X)^2+g^2}\ ,$$
where $k$ is the amplitude, $g$ the resonance width, and $E_0$ the resonance energy. 
The obtained resonance shape shown in Fig.~\ref{fig:concentration}(b) bares close resemblance to the imaginary susceptibility derived from a Kramers-Kronig transformation used in the main text. The absorption coefficient function
$$\alpha(\lambda)=\frac{4\pi\epsilon_{\rm Im}(\lambda)}{2n\lambda}$$
is then used in order to calculate the total absorption of our laser while considering its overlap with the resonance. Finally, $\gamma=(6.3\pm0.5)\%$ is used to calculate the density of photogenerated carriers in the reflection measurement. In Fig.~\ref{fig:concentration}(d) we show the exciton resonance as a function of carrier density. By fitting a linear function $\Delta E = \beta n$ to the data in the low excitation regime we retrieve $\beta=(0.9\pm0.7)\times 10^{-12}$~meV\,cm$^2$ which is a value equivalent to what can be found in other works for similar MoSe$_2$ heterostructures and falls within an order of magnitude to a theoretical estimation of $\beta=5 \times 10^{-12}$~meV\,cm$^2$~\cite{ScuriPRL18, ShahnazaryanPRB1998}.
Finally, by considering the blueshift in the pump-probe measurements we can translate the measured pumping power into the density of photogenerated excitons (see Fig.~5 in the main text). For the data presented in Fig.~5 an average pumping power of  100~\textmu W corresponded to $n=10^{12}$/cm$^2$.

%%%%%%%%%%%%%
\section{Determination of the pulse duration}
\begin{figure}[h!]
	\centering
	\includegraphics[width=0.75\columnwidth]{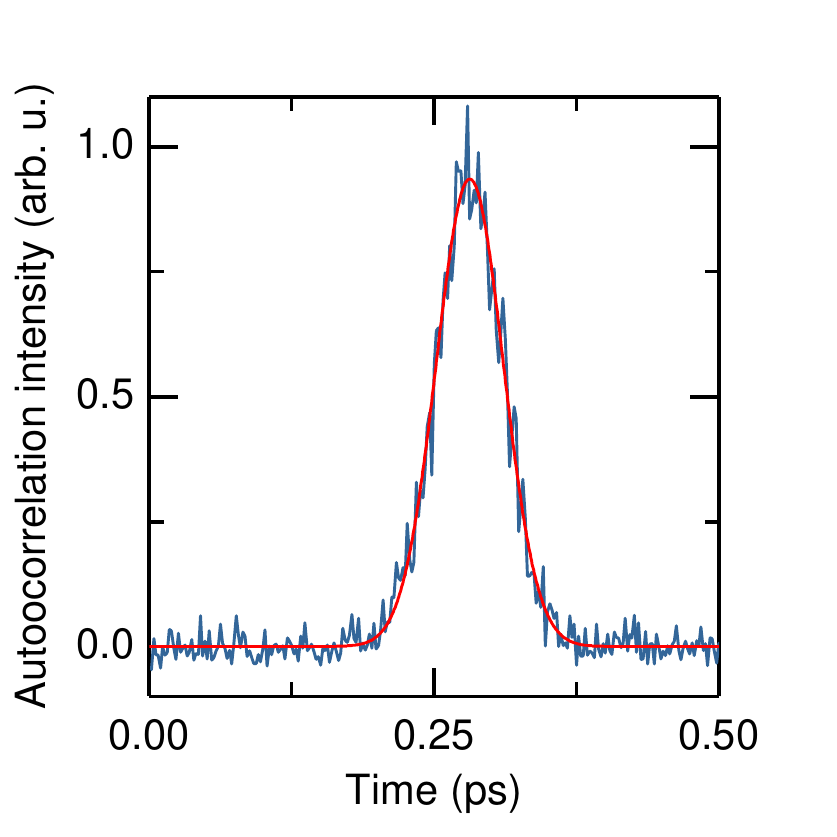}
	\caption{Autocorrelation intensity of the laser beams used in the pump-probe experiments in blue and the fitted Gaussian in red.}
	\label{fig:autocorr}
\end{figure}

The temporal resolution of the pump-probe experiment is given by the laser duration, which we determine by an autocorrelation measurement. In order to take into account possible dispersion effects the probing point for the autocorrelation was chosen right before entering the cryostat after the beams have passed through all optical elements in the setup. The signal intensity is presented in Fig. \ref{fig:autocorr} and fitted with a gaussian function of standard deviation $\Delta t_{\rm int}=30$~fs. To retrieve the correct pulse duration for the electric field in Eq.~(2) in the main text we have to scale fitted value for the intensity via
$$
\Delta t = \frac{\Delta t_{\rm int}}{\sqrt{2}} \approx 21\ {\rm fs}
$$
and a corresponding full width at half maximum (FWHM) of 50~fs.

\bibliographystyle{unsrtnat}
%\bibliography{pp.bib}